\documentclass[sigplan,screen]{acmart}
\setcopyright{none}
\copyrightyear{2025}
\acmConference[Dafny at POPL 2026]{3rd Workshop on Auto-active Programming and Verification Languages (informal proceedings)}{January 11th, 2026}{Rennes, France}
\usepackage{graphicx} % Required for inserting images
\usepackage{graphicx}
\usepackage{listings}
\usepackage{xparse}
\usepackage{expl3}
\usepackage{hyperref}
\usepackage{xcolor}
\lstdefinelanguage{Dafny}{
  keywords={method, function, var, const, type, datatype, class, while, if, else, return, true, false, ensures, requires, assert, assume, lemma, ghost, predicate, forall, exists, invariant, modifies, import, module},
  sensitive=true,
  comment=[l]{//},
  morecomment=[s]{/*}{*/},
  morestring=[b]",
}

\definecolor{dkblue}{RGB}{0,70,140}
\definecolor{dkgreen}{RGB}{0,120,0}
\definecolor{dkred}{RGB}{160,0,0}
\definecolor{gray75}{gray}{0.75}
\definecolor{penblue}{RGB}{0,0,255}

\lstdefinestyle{dafny}{
  language=Dafny,
  lineskip=-1pt,
  aboveskip=2pt,
  belowskip=2pt,
  basicstyle=\ttfamily\normalsize,
  keywordstyle=\bfseries\color{dkgreen},
  commentstyle=\itshape\color{dkgreen},
  stringstyle=\color{purple},
  numberstyle=\tiny\color{gray75},
  numbersep=8pt,
  rulecolor=\color{gray75},
  showstringspaces=false,
  tabsize=2,
  columns=fullflexible,
  breaklines=true,
  breakatwhitespace=true,
  keywordstyle=[2]\color{penblue}, % function and method names
  keywordstyle=[3]\bfseries\color{dkred} % type names
}
\makeatletter
\newcommand*\IfEmptyTF[1]{%
  \edef\@tempa{#1}%
  \ifx\@tempa\@empty
    \expandafter\@firstoftwo
  \else
    \expandafter\@secondoftwo
  \fi
}
\makeatother

\lstnewenvironment{dafny}[3]{
  \lstset{style=dafny,#1}
  \IfEmptyTF{#2}{}{\lstset{morekeywords={[2]#2}}}
  \IfEmptyTF{#3}{}{\lstset{morekeywords={[3]#3}}}
}{}
\newcommand{\dafnyinline}[3]{\lstinline[
    style=dafny,
    #1,
    morekeywords={[2]{#2}},
    morekeywords={[3]{#3}}
]}

\usepackage{fancyvrb}
\usepackage{color}
\usepackage[utf8]{inputenc}
\usepackage{subcaption}  % subcaption is modern and works well
\usepackage{mwe}         % for example-image placeholder; remove for real images
\usepackage{booktabs}

\makeatletter
\def\PY@reset{\let\PY@it=\relax \let\PY@bf=\relax%
    \let\PY@ul=\relax \let\PY@tc=\relax%
    \let\PY@bc=\relax \let\PY@ff=\relax}
\def\PY@tok#1{\csname PY@tok@#1\endcsname}
\def\PY@toks#1+{\ifx\relax#1\empty\else%
    \PY@tok{#1}\expandafter\PY@toks\fi}
\def\PY@do#1{\PY@bc{\PY@tc{\PY@ul{%
    \PY@it{\PY@bf{\PY@ff{#1}}}}}}}
\def\PY#1#2{\PY@reset\PY@toks#1+\relax+\PY@do{#2}}

\@namedef{PY@tok@w}{\def\PY@tc##1{\textcolor[rgb]{0.73,0.73,0.73}{##1}}}
\@namedef{PY@tok@c}{\let\PY@it=\textit\def\PY@tc##1{\textcolor[rgb]{0.24,0.48,0.48}{##1}}}
\@namedef{PY@tok@cp}{\def\PY@tc##1{\textcolor[rgb]{0.61,0.40,0.00}{##1}}}
\@namedef{PY@tok@k}{\let\PY@bf=\textbf\def\PY@tc##1{\textcolor[rgb]{0.00,0.50,0.00}{##1}}}
\@namedef{PY@tok@kp}{\def\PY@tc##1{\textcolor[rgb]{0.00,0.50,0.00}{##1}}}
\@namedef{PY@tok@kt}{\def\PY@tc##1{\textcolor[rgb]{0.69,0.00,0.25}{##1}}}
\@namedef{PY@tok@o}{\def\PY@tc##1{\textcolor[rgb]{0.40,0.40,0.40}{##1}}}
\@namedef{PY@tok@ow}{\let\PY@bf=\textbf\def\PY@tc##1{\textcolor[rgb]{0.67,0.13,1.00}{##1}}}
\@namedef{PY@tok@nb}{\def\PY@tc##1{\textcolor[rgb]{0.00,0.50,0.00}{##1}}}
\@namedef{PY@tok@nf}{\def\PY@tc##1{\textcolor[rgb]{0.00,0.00,1.00}{##1}}}
\@namedef{PY@tok@nc}{\let\PY@bf=\textbf\def\PY@tc##1{\textcolor[rgb]{0.00,0.00,1.00}{##1}}}
\@namedef{PY@tok@nn}{\let\PY@bf=\textbf\def\PY@tc##1{\textcolor[rgb]{0.00,0.00,1.00}{##1}}}
\@namedef{PY@tok@ne}{\let\PY@bf=\textbf\def\PY@tc##1{\textcolor[rgb]{0.80,0.25,0.22}{##1}}}
\@namedef{PY@tok@nv}{\def\PY@tc##1{\textcolor[rgb]{0.10,0.09,0.49}{##1}}}
\@namedef{PY@tok@no}{\def\PY@tc##1{\textcolor[rgb]{0.53,0.00,0.00}{##1}}}
\@namedef{PY@tok@nl}{\def\PY@tc##1{\textcolor[rgb]{0.46,0.46,0.00}{##1}}}
\@namedef{PY@tok@ni}{\let\PY@bf=\textbf\def\PY@tc##1{\textcolor[rgb]{0.44,0.44,0.44}{##1}}}
\@namedef{PY@tok@na}{\def\PY@tc##1{\textcolor[rgb]{0.41,0.47,0.13}{##1}}}
\@namedef{PY@tok@nt}{\let\PY@bf=\textbf\def\PY@tc##1{\textcolor[rgb]{0.00,0.50,0.00}{##1}}}
\@namedef{PY@tok@nd}{\def\PY@tc##1{\textcolor[rgb]{0.67,0.13,1.00}{##1}}}
\@namedef{PY@tok@s}{\def\PY@tc##1{\textcolor[rgb]{0.73,0.13,0.13}{##1}}}
\@namedef{PY@tok@sd}{\let\PY@it=\textit\def\PY@tc##1{\textcolor[rgb]{0.73,0.13,0.13}{##1}}}
\@namedef{PY@tok@si}{\let\PY@bf=\textbf\def\PY@tc##1{\textcolor[rgb]{0.64,0.35,0.47}{##1}}}
\@namedef{PY@tok@se}{\let\PY@bf=\textbf\def\PY@tc##1{\textcolor[rgb]{0.67,0.36,0.12}{##1}}}
\@namedef{PY@tok@sr}{\def\PY@tc##1{\textcolor[rgb]{0.64,0.35,0.47}{##1}}}
\@namedef{PY@tok@ss}{\def\PY@tc##1{\textcolor[rgb]{0.10,0.09,0.49}{##1}}}
\@namedef{PY@tok@sx}{\def\PY@tc##1{\textcolor[rgb]{0.00,0.50,0.00}{##1}}}
\@namedef{PY@tok@m}{\def\PY@tc##1{\textcolor[rgb]{0.40,0.40,0.40}{##1}}}
\@namedef{PY@tok@gh}{\let\PY@bf=\textbf\def\PY@tc##1{\textcolor[rgb]{0.00,0.00,0.50}{##1}}}
\@namedef{PY@tok@gu}{\let\PY@bf=\textbf\def\PY@tc##1{\textcolor[rgb]{0.50,0.00,0.50}{##1}}}
\@namedef{PY@tok@gd}{\def\PY@tc##1{\textcolor[rgb]{0.63,0.00,0.00}{##1}}}
\@namedef{PY@tok@gi}{\def\PY@tc##1{\textcolor[rgb]{0.00,0.52,0.00}{##1}}}
\@namedef{PY@tok@gr}{\def\PY@tc##1{\textcolor[rgb]{0.89,0.00,0.00}{##1}}}
\@namedef{PY@tok@ge}{\let\PY@it=\textit}
\@namedef{PY@tok@gs}{\let\PY@bf=\textbf}
\@namedef{PY@tok@ges}{\let\PY@bf=\textbf\let\PY@it=\textit}
\@namedef{PY@tok@gp}{\let\PY@bf=\textbf\def\PY@tc##1{\textcolor[rgb]{0.00,0.00,0.50}{##1}}}
\@namedef{PY@tok@go}{\def\PY@tc##1{\textcolor[rgb]{0.44,0.44,0.44}{##1}}}
\@namedef{PY@tok@gt}{\def\PY@tc##1{\textcolor[rgb]{0.00,0.27,0.87}{##1}}}
\@namedef{PY@tok@err}{\def\PY@bc##1{{\setlength{\fboxsep}{\string -\fboxrule}\fcolorbox[rgb]{1.00,0.00,0.00}{1,1,1}{\strut ##1}}}}
\@namedef{PY@tok@kc}{\let\PY@bf=\textbf\def\PY@tc##1{\textcolor[rgb]{0.00,0.50,0.00}{##1}}}
\@namedef{PY@tok@kd}{\let\PY@bf=\textbf\def\PY@tc##1{\textcolor[rgb]{0.00,0.50,0.00}{##1}}}
\@namedef{PY@tok@kn}{\let\PY@bf=\textbf\def\PY@tc##1{\textcolor[rgb]{0.00,0.50,0.00}{##1}}}
\@namedef{PY@tok@kr}{\let\PY@bf=\textbf\def\PY@tc##1{\textcolor[rgb]{0.00,0.50,0.00}{##1}}}
\@namedef{PY@tok@bp}{\def\PY@tc##1{\textcolor[rgb]{0.00,0.50,0.00}{##1}}}
\@namedef{PY@tok@fm}{\def\PY@tc##1{\textcolor[rgb]{0.00,0.00,1.00}{##1}}}
\@namedef{PY@tok@vc}{\def\PY@tc##1{\textcolor[rgb]{0.10,0.09,0.49}{##1}}}
\@namedef{PY@tok@vg}{\def\PY@tc##1{\textcolor[rgb]{0.10,0.09,0.49}{##1}}}
\@namedef{PY@tok@vi}{\def\PY@tc##1{\textcolor[rgb]{0.10,0.09,0.49}{##1}}}
\@namedef{PY@tok@vm}{\def\PY@tc##1{\textcolor[rgb]{0.10,0.09,0.49}{##1}}}
\@namedef{PY@tok@sa}{\def\PY@tc##1{\textcolor[rgb]{0.73,0.13,0.13}{##1}}}
\@namedef{PY@tok@sb}{\def\PY@tc##1{\textcolor[rgb]{0.73,0.13,0.13}{##1}}}
\@namedef{PY@tok@sc}{\def\PY@tc##1{\textcolor[rgb]{0.73,0.13,0.13}{##1}}}
\@namedef{PY@tok@dl}{\def\PY@tc##1{\textcolor[rgb]{0.73,0.13,0.13}{##1}}}
\@namedef{PY@tok@s2}{\def\PY@tc##1{\textcolor[rgb]{0.73,0.13,0.13}{##1}}}
\@namedef{PY@tok@sh}{\def\PY@tc##1{\textcolor[rgb]{0.73,0.13,0.13}{##1}}}
\@namedef{PY@tok@s1}{\def\PY@tc##1{\textcolor[rgb]{0.73,0.13,0.13}{##1}}}
\@namedef{PY@tok@mb}{\def\PY@tc##1{\textcolor[rgb]{0.40,0.40,0.40}{##1}}}
\@namedef{PY@tok@mf}{\def\PY@tc##1{\textcolor[rgb]{0.40,0.40,0.40}{##1}}}
\@namedef{PY@tok@mh}{\def\PY@tc##1{\textcolor[rgb]{0.40,0.40,0.40}{##1}}}
\@namedef{PY@tok@mi}{\def\PY@tc##1{\textcolor[rgb]{0.40,0.40,0.40}{##1}}}
\@namedef{PY@tok@il}{\def\PY@tc##1{\textcolor[rgb]{0.40,0.40,0.40}{##1}}}
\@namedef{PY@tok@mo}{\def\PY@tc##1{\textcolor[rgb]{0.40,0.40,0.40}{##1}}}
\@namedef{PY@tok@ch}{\let\PY@it=\textit\def\PY@tc##1{\textcolor[rgb]{0.24,0.48,0.48}{##1}}}
\@namedef{PY@tok@cm}{\let\PY@it=\textit\def\PY@tc##1{\textcolor[rgb]{0.24,0.48,0.48}{##1}}}
\@namedef{PY@tok@cpf}{\let\PY@it=\textit\def\PY@tc##1{\textcolor[rgb]{0.24,0.48,0.48}{##1}}}
\@namedef{PY@tok@c1}{\let\PY@it=\textit\def\PY@tc##1{\textcolor[rgb]{0.24,0.48,0.48}{##1}}}
\@namedef{PY@tok@cs}{\let\PY@it=\textit\def\PY@tc##1{\textcolor[rgb]{0.24,0.48,0.48}{##1}}}

\makeatother
\usepackage{tikz}
\usetikzlibrary{positioning, arrows.meta}

\title{The Design of an Interactive Proof Mode for Dafny}

\author{Ștefan Ciobâcă}
\email{stefan.ciobaca@uaic.ro}
\orcid{https://orcid.org/0009-0000-4082-570X}
\affiliation{%
  \institution{Alexandru Ioan Cuza University}
  \city{Iași}
  \country{Romania}
}
\authornote{Authors are listed in alphabetical order.}

\author{K. Rustan M. Leino}
\email{leino@amazon.com}
\orcid{https://orcid.org/0000-0003-2872-8039}
\affiliation{%
  \institution{Amazon Web Services Inc}
  \city{Seattle}
  \state{WA}
  \country{US}
}
\authornotemark[1]

\author{Ștefan-Alexandru Mercaș}
\email{smercas@gmail.com}
\orcid{https://orcid.org/0009-0001-1265-0348}
\affiliation{%
  \institution{Alexandru Ioan Cuza University}
  \city{Iași}
  \country{Romania}
}
\authornotemark[1]

\author{Roxana-Mihaela Timon}
\email{timonroxana91@gmail.com}
\orcid{https://orcid.org/0009-0005-9867-9633}
\affiliation{%
  \institution{Alexandru Ioan Cuza University}
  \city{Iași}
  \country{Romania}
}
\authornotemark[1]

\begin{abstract}
We propose to extend the \texttt{Dafny} system with an interactive proof mode. We present a motivating example, how the IPM works, including the main design choices we make, and a prototype implementation.
\end{abstract}

\begin{document}




\maketitle

\section{Introduction}

\texttt{Dafny}~\cite{DBLP:conf/lpar/Leino10} is a programming language that supports auto-active verification. \texttt{Dafny} works by translation into the intermediate verification language \texttt{Boogie}~\cite{boogie}, which in turn discharges verification obligations using an SMT solver, typically \texttt{Z3}~\cite{z3}.

Due to the inherent computational complexity of the underlying problem, proof obligations are in general not discharged automatically. Instead, the programmer must guide the prover by providing helper proof annotations. These annotations could consist of helper assertions, loop invariants, read/write frames, helper lemmas, ghost state/code, and others. Once a proof annotation is added to the code, the entire development needs to be recompiled in order to test whether the annotation truly helps the underlying solver. 

\begin{figure}[t!]
\centering
\begin{dafny}{}{triangle_sum_even}{int}
lemma triangle_sum_even(x : int)
  ensures x * (x + 1) % 2 == 0
{
  assume false;
}
\end{dafny}
\caption{A \texttt{Dafny} lemma that fails to verify automatically, which we use as a running example.}
\label{fig:lemma}
\hrulefill
\Description[Dafny code showing a lemma that does not verify automatically.]{Dafny code showing a lemma that does not verify automatically. The lemma states that the sum of the first x naturals is even.}
\end{figure}

We propose an Interactive Proof Mode (IPM) for \texttt{Dafny}. The IPM starts with a proof obligation that cannot be discharged automatically by the \texttt{Z3} solver. It prints the proof obligation as a set of Boolean-valued \texttt{Dafny} expressions that represent \( \bullet \) the hypotheses and \( \bullet \) the goal to be proven from the hypotheses. It accepts input from the user in the form of tactics to be applied on the current proof obligation, similar to proof assistants, such as \texttt{Lean}~\cite{leanReferenceManual} and \texttt{Rocq}~\cite{rocqReferenceManual}. Once the proof obligation is discharged, the IPM prints out \texttt{Dafny} code that represents a formal proof of the original verification condition. The proof can be copied back into the \texttt{Dafny} development at the point where the initial proof obligation occurs in the code, serving as an explanation for why the assertion that could not be discharged automatically holds. Unless there is a bug in the IPM, the code now has enough proof annotations for the proof obligation to discharge successfully.

\section{Motivating Example}

We consider the \texttt{Dafny} lemma in Figure~\ref{fig:lemma}, which cannot be proven automatically.

Because of the use of non-linear integer arithmetic, the proof obligation generated by the lemma is not discharged automatically by \texttt{Z3} when called with the default configuration options set by \texttt{Dafny}.

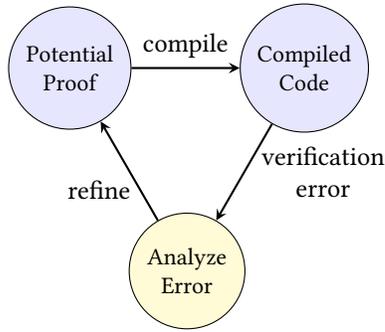
\begin{figure}[ht]
\begin{center}
\begin{tikzpicture}[
  scale=0.6,
  arrow/.style={->, >=stealth, thick},
  state/.style={circle, draw, minimum size=1cm, align=center, fill=blue!10, font=\small},
  lightbulb/.style={circle, draw, minimum size=1cm, align=center, fill=yellow!20, font=\small}
]

\node[state] (potential) at (150:3cm) {Potential\\Proof};
\node[state] (compiled) at (30:3cm) {Compiled\\Code};
\node[lightbulb] (analyze) at (270:3cm) {Analyze\\Error};

\draw[arrow] (potential) -- (compiled) node[midway, above, align=center] {compile};
\draw[arrow] (compiled) -- (analyze) node[midway, right, align=center, xshift=1mm] {verification \\ error};
\draw[arrow] (analyze) -- (potential) node[midway, below, align=center, xshift=-4mm] {refine};
\end{tikzpicture}
\end{center}
\caption{Typical Workflow in \texttt{Dafny} Development}
\label{fig:typical-workflow}
\hrulefill
\Description[The typical workflow for developing Dafny programs.]{The typical workflow for developing Dafny programs. We start with a potential proof and if there is a verification error, we analyze it and refine the proof.}
\end{figure}

When confronted with a verification obligation that fails, a \texttt{Dafny} programmer will typically start with the placeholder proof \verb|assume false;|, and enter an iterative refinement process to build a proof of the failing proof obligation, shown in Figure~\ref{fig:typical-workflow}.

\begin{figure*}[t!]
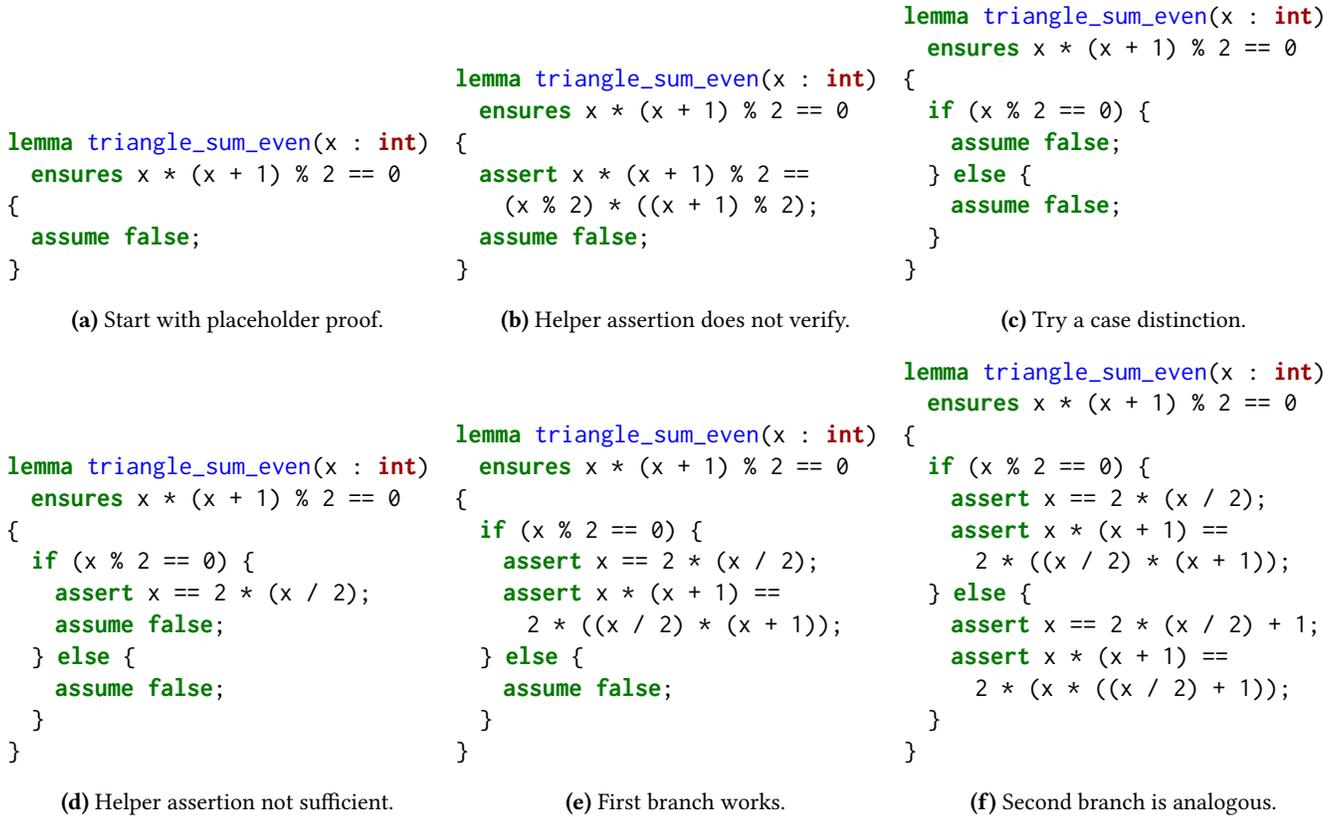
 % figure* spans both columns in two-column formats
  \centering
  % 4 columns across: each subfigure gets 0.24\textwidth (4 * 0.24 = 0.96 < 1.0)
  \begin{subfigure}[b]{0.33\textwidth}
    \centering
\begin{dafny}{}{triangle_sum_even}{int}
lemma triangle_sum_even(x : int)
  ensures x * (x + 1) % 2 == 0
{
  assume false;
}
\end{dafny}
    \caption{Start with placeholder proof.}
    \label{fig:sub1}
  \end{subfigure}\hfill
  \begin{subfigure}[b]{0.33\textwidth}
    \centering
\begin{dafny}{}{triangle_sum_even}{int}
lemma triangle_sum_even(x : int)
  ensures x * (x + 1) % 2 == 0
{
  assert x * (x + 1) % 2 ==
    (x % 2) * ((x + 1) % 2);
  assume false;
}
\end{dafny}
    \caption{Helper assertion does not verify.}
    \label{fig:sub2}
  \end{subfigure}\hfill
  \begin{subfigure}[b]{0.33\textwidth}
    \centering
\begin{dafny}{}{triangle_sum_even}{int}
lemma triangle_sum_even(x : int)
  ensures x * (x + 1) % 2 == 0
{
  if (x % 2 == 0) {
    assume false;
  } else {
    assume false;
  }
}
\end{dafny}
    \caption{Try a case distinction.}
    \label{fig:sub3}
  \end{subfigure}\hfill

  \vspace{6pt} % small vertical gap between rows

  \begin{subfigure}[b]{0.33\textwidth}
    \centering
\begin{dafny}{}{triangle_sum_even}{int}
lemma triangle_sum_even(x : int)
  ensures x * (x + 1) % 2 == 0
{
  if (x % 2 == 0) {
    assert x == 2 * (x / 2);
    assume false;
  } else {
    assume false;
  }
}
\end{dafny}
    \caption{Helper assertion not sufficient.}
    \label{fig:sub5}
  \end{subfigure}\hfill
  \begin{subfigure}[b]{0.33\textwidth}
    \centering
\begin{dafny}{}{triangle_sum_even}{int}
lemma triangle_sum_even(x : int)
  ensures x * (x + 1) % 2 == 0
{
  if (x % 2 == 0) {
    assert x == 2 * (x / 2);
    assert x * (x + 1) ==
      2 * ((x / 2) * (x + 1));
  } else {
    assume false;
  }
}
\end{dafny}
    \caption{First branch works.}
    \label{fig:sub6}
  \end{subfigure}\hfill
  \begin{subfigure}[b]{0.33\textwidth}
    \centering
\begin{dafny}{}{triangle_sum_even}{int}
lemma triangle_sum_even(x : int)
  ensures x * (x + 1) % 2 == 0
{
  if (x % 2 == 0) {
    assert x == 2 * (x / 2);
    assert x * (x + 1) ==
      2 * ((x / 2) * (x + 1));
  } else {
    assert x == 2 * (x / 2) + 1;
    assert x * (x + 1) ==
      2 * (x * ((x / 2) + 1));
  }
}
\end{dafny}
    \caption{Second branch is analogous.}
    \label{fig:sub7}
  \end{subfigure}\hfill

  \caption{The stages that a \texttt{Dafny} programmer could go through to prove the lemma in Figure~\ref{fig:lemma}.}
  \label{fig:iterativeproof}
\hrulefill
\Description[The stages that a Dafny programmer could go through to prove the lemma.]{}
\end{figure*}

This iterative proof development loop described above suffers from at least two issues: \( \bullet \) it can be time-consuming (for large code bases) to recompile the entire development with each iteration of the proof, and \( \bullet \) the facts known to the prover at the point where the proof obligation fails are not all locally available, in general, but they can be spread throughout the code (e.g., in some loop invariant, in some class invariant, as a well-formedness condition).

Figure~\ref{fig:iterativeproof} shows a series of steps a programmer might take to construct a proof for the lemma above.

Our interactive proof mode (IPM) addresses these two issues, by allowing developers to debug in an interactive environment a proof obligation that fails to verify. To enter the interactive proof mode, we add the annotation \texttt{:ipm} to the failing proof obligation:

\begin{dafny}{}{triangle_sum_even}{int}
lemma triangle_sum_even(x : int)
  ensures {:ipm} x * (x + 1) % 2 == 0
{ }
\end{dafny}

\noindent and we call a Python script on the \texttt{Dafny} source file: \texttt{python3 \texttt{Dafny}-ipm.py file.dfy}.

The interactive environment resembles that of the  \texttt{Rocq}~\cite{rocqReferenceManual} or \texttt{Lean}~\cite{leanReferenceManual} provers. The proof context (facts known to the system) and the proof goal (what must be shown) are shown at each iteration:
\begin{Verbatim}
1 goal(s) remaining
current goal:
hypotheses
goal
   (((x * (x + 1)) % 2) == 0)
>
\end{Verbatim}
The system reads a series of proof tactics that help discharge the current proof obligation. For example, the tactic \texttt{check} can be used to check whether \texttt{Z3} can prove some fact from the current set of hypotheses:

\begin{Verbatim}
> check (x * (x + 1) % 2 == 
        (x % 2) * ((x + 1) % 2))
No, Z3 cannot prove (x * (x + 1) % 2 == 
                    (x % 2) * ((x + 1) % 2)).
>
\end{Verbatim}

The user receives immediate feedback after each tactic application, as there is no need to recompile the entire code base. The only sources of lag are calls to the solver (\texttt{Z3}), which cannot be avoided, but the timeout can be set to a low value, like \( 1 \) second. To solve the goal above, the user can perform a case analysis using the \texttt{case} tactic:

\begin{Verbatim}
> case ((x % 2) == 0)
2 goal(s) remaining
current goal:
hypotheses
   ((x % 2) == 0)
goal
   (((x * (x + 1)) % 2) == 0)

goal 2 is: (((x * (x + 1)) % 2) == 0)
\end{Verbatim}

The case analysis tactic above splits the current proof obligation into two cases, depending on the truth value of the expression \verb|((x % 2) == 0)|. For the two cases in our example, the user can provide the proof using the tactic \texttt{assert}, which adds a (provable) expression to the current set of hypotheses. After applying a series of tactics that discharge the proof obligation, the IPM prints out a proof of the verification condition:

\begin{Verbatim}
Congrats, current goal proved.
Goal: (((x * (x + 1)) % 2) == 0)
Proof:
if (((x % 2) == 0)) {
  assert (x == (2 * (x / 2)));
  assert ((x * (x + 1)) == 
         (2 * ((x / 2) * (x + 1))));
} else {
  assert (x == ((2 * (x / 2)) + 1));
  assert ((x * (x + 1)) == 
         (2 * (x * ((x / 2) + 1))));
}
\end{Verbatim}

The proof code can be copied back into the \texttt{Dafny} development, and the proof obligation will now verify successfully.

\section{How it works}

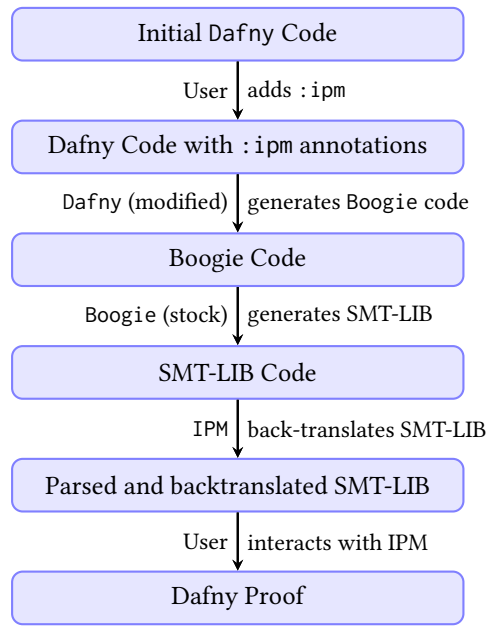
\begin{figure}[t!]
\centering
\begin{tikzpicture}[
  box/.style={rectangle, rounded corners, draw=blue!50, fill=blue!10, 
              thick, minimum width=6cm, minimum height=0.7cm, align=center},
  arrow/.style={->, >=stealth, thick},
  label/.style={font=\small, midway}
]

% Boxes
\node[box] (dafny) at (0,0) {Initial \texttt{Dafny} Code};
\node[box] (ipm) at (0,-1.5) {Dafny Code with \texttt{:ipm} annotations};
\node[box] (boogie) at (0,-3) {Boogie Code};
\node[box] (smtlib) at (0,-4.5) {SMT-LIB Code};
\node[box] (parsed) at (0,-6) {Parsed and backtranslated SMT-LIB};
%\node[box] (interaction) at (0,-7.5) {Interactive Proof Session};
\node[box] (final) at (0,-7.5) {Dafny Proof};

% Arrows with labels
\draw[arrow] (dafny) -- (ipm) node[label, right] {adds \texttt{:ipm}};
\draw[arrow] (dafny) -- (ipm) node[label, left] {User};
\draw[arrow] (ipm) -- (boogie) node[label, right] {generates \texttt{Boogie} code};
\draw[arrow] (ipm) -- (boogie) node[label, left] {\texttt{Dafny} (modified)};
\draw[arrow] (boogie) -- (smtlib) node[label, right] {generates SMT-LIB};
\draw[arrow] (boogie) -- (smtlib) node[label, left] {\texttt{Boogie} (stock)};
\draw[arrow] (smtlib) -- (parsed) node[label, right] {back-translates SMT-LIB};
\draw[arrow] (smtlib) -- (parsed) node[label, left] {\texttt{IPM}};
\draw[arrow] (parsed) -- (final) node[label, right] {interacts with IPM};
\draw[arrow] (parsed) -- (final) node[label, left] {User};
%\draw[arrow] (interaction) -- (final) node[label, right, xshift=2mm] {copy proof back};

\end{tikzpicture}
\caption{IPM Workflow}
\label{fig:ipm-workflow}
\hrulefill
\Description[The workflow of the interactive proof mode that we propose]{We annotate the initial Dafny code with ipm annotations for the proof obligations that we want to debug, we run Dafny-IPM to generate the Boogie and SMT-LIB output, we parse and backtranslate the SMT-LIB file, and we finally output the Dafny proof.}
\end{figure}

The diagram in Figure~\ref{fig:ipm-workflow} shows the internal workshop in the interactive proof mode, which we describe in detail in this section.

The user first annotates the proof obligation for which the interactive proof mode should be started with the \dafnyinline{}{}{}!{:ipm}! attribute (for interactive proof mode). This annotation is currently supported for assertions (i.e., \dafnyinline{}{}{}!assert {:ipm} ...;!) and \texttt{ensures} clauses (i.e., \dafnyinline{}{}{}!ensures {:ipm} ...!). In the future, we plan to also support invariants and other proof obligations.

\paragraph{Instrumentation} We run a modified version of \texttt{Dafny} called \texttt{Dafny-IPM} on the source code\footnote{In future work, we plan to integrate the changes in \texttt{Dafny-IPM} into \texttt{Dafny} and expose them as a command-line argument.}. \texttt{Dafny-IPM} serves two purposes: \( \bullet \) it labels the verification obligations that should be proven in the IPM, and \( \bullet \) it instruments the \texttt{Dafny} expressions in the source code in order to make it easier to back-translate from SMT-LIB to \texttt{Dafny}.

To this end, \texttt{Dafny-IPM} programmatically adds three ghost functions to each module, called \dafnyinline{}{_protect}{}!_protect!, \dafnyinline{}{_protectScope}{}!_protectScope!, and \dafnyinline{}{_protectToProve}{}!_protectToProve!, which are defined as follows:

\begin{dafny}{breaklines=false}{_protect,_protectScope,_protectToProve}{T,string,seq,bool}
function _protect<T>(x: T, name: string): T
  { x }
function _protectScope<T>(
  x: T, name: string): bool { true }
function _protectToProve<T>(
  x: T, name: string, scope: seq<bool>): T { x }
\end{dafny}

We use names starting with \texttt{\_} for the three functions because the \texttt{Dafny} parser does not accept such identifiers, but they can be used successfully in subsequent passes. Therefore they effectively act as reserved names and cannot clash with user-defined identifiers.

The function \dafnyinline{}{_protectToProve}{}!_protectToProve! is used as a box that identifies which proof obligations should go to the IPM. The function \dafnyinline{}{_protect}{}!_protect! is used to box each occurrence of a \texttt{Dafny} variable in order to back-translate SMT-LIB identifiers, as we explain later in the paper in more detail. Finally, the function \dafnyinline{}{_protectScope}{}!_protectScope! is used to help the IPM resolve the variables currently in scope in the cases where name shadowing occurs, as we explained below.

For example, consider the following \texttt{Dafny} code:

\begin{dafny}{}{Example}{int}
method Example(x : nat, y : nat)
  requires x + y > 0
  ensures x + y > 0
{
  var x : nat := 1;
  assert {:ipm} x + y > 0;
}
\end{dafny}

Dafny-IPM instruments the code above as follows:

\begin{dafny}{breaklines=false}{Example,_protect,_protectToProve,_protectScope}{int}
method Example(x : nat, y : nat)
  requires _protect(x,"x") + _protect(y,"y") > 0
  ensures _protect(x,"x") + _protect(y,"y") > 0
{
  var x : nat := 1;
  assert {:ipm} _protectToProve(
    _protect(x, "x") > _protect(y, "y"), 
    "x + y > 0", 
    [_protectScope(x,"x"),
     _protectScope(y,"y")]);
}
\end{dafny}

% \texttt{Dafny} allows for name shadowing: a variable can be declared with an identifier used to refer to another variable that was defined in one of the enclosing scopes, thus shadowing it. With few exceptions (such as class variables), the shadowed variables cannot be referred to within the scope of the shadowing variable after its declaration. In order to determine the scopes in relevant contexts (and thus solving any ambiguity that might appear due to variable shadowing), we employ the help of the \texttt{Dafny} resolver to fill out the scope sequence of the third argument passed to \dafnyinline{}{_protectToProve}{}!_protectToProve!.

% For example, consider the following code:

%\begin{dafny}{}{}{real,nat,char,this}
%class C {
  %var n: real
  %method main(n: nat) {
    %var n: char := 'n';
    %assert this.n is real;
    %assert n == 'n';
  %}
%}
%\end{dafny}

%In this example, the first two variables with the identifier of "\dafnyinline{}{}{}!n!" are shadowed within the contexts of the assertions. However, the class field remains accessible and can still be explicitly referenced using "\dafnyinline{}{}{this}!this.n!".

The IPM then generates the \texttt{Boogie} code using the \texttt{-print} command-line option and calls stock \texttt{Boogie} on the resulting file to generate the verification conditions in SMT-LIB format.

\paragraph{Back-translation of SMT-LIB} The IPM then parses the SMT-LIB code using the \texttt{Python} \texttt{Z3} API. The translation from SMT-LIB to \texttt{Dafny} constitutes the central component of our interactive verification system.

Here is an extract of the SMT-LIB file generated for the lemma in Figure~\ref{fig:lemma} by stock \texttt{Dafny}:

\begin{Verbatim}[fontsize=\small]
[...]
(assert (forall ((o@@5 T@Box) ) (!  
 (not (|Set#IsMember| |Set#Empty| o@@5))
 :qid |filebpl.796:15|
 :skolemid |125|
 :pattern ( (|Set#IsMember| |Set#Empty| o@@5))
)))
[...]
(push 1)
[...]
(assert (not (=> (= (ControlFlow 0 0) 2) (let
 ((anon0_correct (=> (and (and ($IsGoodHeap $Heap)
 ($IsHeapAnchor $Heap)) (and (= $_ModifiesFrame@0
 (|lambda#0| null $Heap alloc false)) (= (ControlFlow
 0 2) (- 0 1)))) (= (Mod (Mul |x#0@@1| (+ |x#0@@1| 1))
 (LitInt 2)) (LitInt 0))))) anon0_correct)) ))
(check-sat)
[...]
(pop 1)
\end{Verbatim}

The file contains some solver configuration options (omitted for brevity), around \( 1000 \) axioms (an example axiom about sets shown above) and then the verification obligations, each verification obligation surrounded by \verb|(push 1)|/\verb|(pop 1)|. The core of the verification obligation is 

\begin{Verbatim}[fontsize=\small]
(= (Mod (Mul |x#0@@1| (+ |x#0@@1| 1)) (LitInt 2)) 
   (LitInt 0)),
\end{Verbatim}

\noindent which corresponds to the \texttt{Dafny} post-condition of the lemma, \texttt{x * (x + 1) \% 2 == 0}.

Note that: \( \bullet \) the identifier \texttt{x} in the initial source code is now mangled: it appears as \texttt{x\#0@@1} in SMT-LIB, that \( \bullet \) the arithmetic operations are symbols like \texttt{Mod} and \texttt{Mul}, which are axiomatized, that \( \bullet \) the constants are boxed (e.g., \texttt{LitInt 2} for the constant \texttt{2}), and that \( \bullet \) additional constructions, like \texttt{\$IsGoodHeap}, are not present in the original source code, but are an artifact of the translation process.

In order to be able to back-translate the verification condition into \texttt{Dafny}, we work not with the SMT-LIB file generated by stock \texttt{Dafny}, but with the SMT-LIB file corresponding to the instrumented \texttt{Dafny} code.

In the instrumented version, the identifiers in the verification condition are additionally wrapped in a call to \texttt{\_protect}. For example, all occurrences of \texttt{|x\#0@@1|} seen in the stock verification obligation above are replaced by the following expression:

\begin{Verbatim}[fontsize=\small]
($Unbox_995 (_module.__default.__protect TInt 
reveal__module._default._protect 
($Box_577 |x#0@@1|) (Lit_25231 (|Seq#Build| 
|Seq#Empty| ($Box_23439 (|char#FromInt| 120))))))
\end{Verbatim}

This represents a call to \texttt{\_protect} with the arguments \texttt{x} (in SMT-LIB: \texttt{|x\#0@@1|}) and \texttt{"x"} (in SMT-LIB: \texttt{[...] \texttt{120} [...]}, where \( 120 \) is the numerical code for lowercase \texttt{x}). Although the verification condition becomes even larger in this way, and therefore more difficult to understand by a human, our processing framework can back-translate it into \texttt{Dafny} mechanically.
%
%In addition, the verification obligations that are labeled by \texttt{\{:ipm\}} appear under a call to \texttt{\_protectToProve}. In the other proof obligations, there is no call to \texttt{\_protectToProve}.

Using string processs, we first split the SMT-LIB file into axioms and verification conditions, and we parse each using the Python \texttt{Z3} API\footnote{We cannot parse the entire file directly, because the parser provided by the \texttt{Z3} API ignores commands like \texttt{(push 1)} or setting configuration options.}. We extract from the proof obligation of the form \( \varphi_1 \rightarrow (\varphi_2 \ldots (\varphi_n \rightarrow \varphi))) \) the hypotheses \( \varphi_1, \ldots \varphi_n \) and the goal \( \varphi \). 

We back-translate the hypotheses and the conclusion as follows:

\begin{enumerate}

\item we replace symbols, such as \texttt{Mod} and \texttt{Mul}, with their mathematical counterparts;

\item we unbox integer literals;

\item we remove all calls to \texttt{\_protect}, and simultaneously store data structures that map \texttt{Dafny} identifiers to SMT-LIB identifiers (e.g., \texttt{x} to \texttt{|x\#0@@1|}) and vice-versa;

\item when a \texttt{Dafny} name is shadowed, like the identifier \texttt{x} in the  method \dafnyinline{}{Example}{}!Example! above, it corresponds to two SMT-LIB names (like \texttt{|x\#1@@1|} and \texttt{|x\#0@@1|}), and we prioritize the SMT-LIB identifier that occurs under \dafnyinline{}{_protectScope}{}!_protectScope!, which corresponds to the \texttt{Dafny} identifier that is visible at the point of the assertion being proved interactively;

\item we remove all instances of \texttt{ControlFlow}, which is only used by Boogie to report locations of errors;

\item we (soundly) remove calls to various functions, such as \texttt{\$IsGoodHeap} and \texttt{\$IsHeapAnchor}, which are implicit in the \texttt{Dafny} code;

\item we (currently) remove calls to \texttt{\$\_ModifiesFrame}, calls that we plan to handle in the future, when we add support for the imperative parts;

\item we simplify expressions by removing hypotheses of the form \( \textit{name} = \textit{expression} \) if \textit{name} does not occur anywhere else (these can occur due to \textit{name} being used in parts of the expressions that we ignore, like \texttt{\$IsGoodHeap});

\item we identify which verification obligations should be used in the IPM by checking whether there is a call to \texttt{\_protectToProve} in the goal.

\end{enumerate}

In order for the IPM to be faithful to the \texttt{Dafny} pipeline, the operations above are only applied for the purpose of pretty-printing (with the exception of removing the calls to \texttt{\_protect*}). The original proof obligation is kept in memory and is used to query the solver during the interactive proof session.

\subsection{Tactics}

The IPM provides a set of tactics that allow users to guide the proof process. These tactics transform the initial proof obligation into simpler proof obligations.

We currently support the following tactics:

\begin{enumerate}
    \item \texttt{check};
    \item \texttt{assert};
    \item \texttt{case};
    \item \texttt{assume}.
\end{enumerate}

\( \bullet \) The tactic \texttt{check} takes a \texttt{Dafny} boolean-valued expression \( \psi \) as argument and checks whether \( \psi \) is a consequence of the current hypotheses. It does not update the proof obligation. It is useful to understand whether the prover can prove some fact. To perform the check, we assume that the current proof obligation is of the form \( \varphi_1 \rightarrow (\varphi_2 \rightarrow \ldots (\varphi_n \rightarrow \varphi)) \). The goal \( \varphi \) is replaced by \( \psi \) and the new verification condition is sent to the solver to check for its validity. The IPM reports the response of the solver. We currently parse the expression \( \psi \) using the Python \texttt{ast} module. Each node of the resulting Python AST is then translated into a \texttt{Z3} expression by mapping Python language constructs to their \texttt{Z3} counterparts. Names and constants (\texttt{ast.Name}, \texttt{ast.Constant}) are mapped to \texttt{Z3} variables and values. The final result is a \texttt{Z3} formula. This works because the sub-language of expressions that we currently support in the IPM, which includes arithmetic operations, relational operations, boolean operators, and conditional expressions is close enough to Python. 

\( \bullet \) The \texttt{assert} tactic introduces an intermediate assertion that decomposes the proof into two sub-objectives: \textbf{proving the assertion} (the assertion follows from the current hypotheses)
and \textbf{using the assertion}: the original goal follows from the original hypotheses, to which we add the helper assertion.

Formally, the tactic \texttt{assert} \( \psi \) transforms the current proof obligation, \( \varphi_1 \rightarrow (\varphi_2 \rightarrow \ldots (\varphi_n \rightarrow \varphi)) \), into the following two proof obligations: \( \bullet \) \( \varphi_1 \rightarrow (\varphi_2 \rightarrow \ldots (\varphi_n \rightarrow \psi)) \) and \( \bullet \) \( \varphi_1 \rightarrow (\varphi_2 \rightarrow \ldots (\varphi_n \rightarrow (\psi \rightarrow \varphi))) \).

The proof obligations that cannot be discharged automatically by the solver are presented to the user, who must provide additional tactics.

\( \bullet \) The \texttt{case} tactic implements case analysis, depending on the truth value of the argument.

Formally, the tactic \texttt{case} \( \psi \) transforms the current proof obligation, \( \varphi_1 \rightarrow (\varphi_2 \rightarrow \ldots (\varphi_n \rightarrow \varphi)) \), into the following two proof obligations: \( \bullet \) \( \varphi_1 \rightarrow (\varphi_2 \rightarrow \ldots (\varphi_n \rightarrow (\psi \rightarrow \varphi))) \) and \( \bullet \) \( \varphi_1 \rightarrow (\varphi_2 \rightarrow \ldots (\varphi_n \rightarrow (\lnot \psi \rightarrow \varphi))) \).

\( \bullet \) The \texttt{assume} tactic supports the introduction of an additional hypothesis without proof. It is helpful to experiment with questions of the form: would adding this hypothesis make the proof go through?

\paragraph{Solver Interaction} Before a proof obligation is shown to the user by the IPM, the system sends it to the solver. Only the proof obligations that are not handled automatically are shown to the user. In order to be faithful to the \texttt{Dafny/Boogie} verification pipeline, we manually set the \texttt{Z3} configurations to those that are set by default by the stock \texttt{Dafny} pipeline: \texttt{mbqi} to false, solver runs in incremental mode, etc.

\paragraph{Other commands} The system also supports for an \texttt{undo} operation. This is handled by storing a history of all proof obligations after each tactic application. The other command that we allow are \texttt{:help}, which shows the available commands, and \texttt{:quit}, which abandons the proof effort and exits the IPM. In the future, we plan to add other commands, such as for searching axioms, switching among proof obligations, etc.

Each proof obligation stores the tactic that the user entered to solve it, and pointers to the resulting proof obligations. In this way, once all proof obligations are discharged, the tactics are used to reconstruct a \texttt{Dafny} proof, which is shown to the user. The proof can be copied back into the initial source code as a proof/explanation of the proof obligation that was initially failing.

\section{Related Work}

Many other auto-active verification systems~\cite{why3ReferenceManual, fstarReferenceManual, 10.1007/s10817-022-09619-1, McCormick_Chapin_2015} rely on helper assertions to guide the underlying solver. Out of these tools, the Why3 deductive verifier~\cite{why3ReferenceManual} typically relies on annotations (invariants, assertions), but also supports for interactive proofs. The interactive mode of Why3 is much more mature than our prototype for \texttt{Dafny}. However, unlike the proof mode that we propose, the proof object that is obtained in Why3 using the interactive mode is hidden, and cannot be placed into  the initial source code to serve as an explanation of why the verification obligation holds. Another approach to the same problem of integrating automated and interactive provers is to add automation tactics into interactive theorem provers~\cite{SledgehammerGuide}. Recent work~\cite{10.1145/3632911} shows how to combine interactive and automated proofs for \texttt{C} programs in the \texttt{Rocq} prover. A monadic shallow embedding of an executable program semantics into \texttt{Lean} is used to obtain a framework~\cite{gladshtein2026foundational} where \texttt{Dafny}-style specifications can be discharged using a combination of off-the-shelf SMT solvers and interactive proofs. An embedding~\cite{rozowski2025leanondafny} of a fragment of \texttt{Dafny} into \texttt{Lean} with the goal of allowing for interactive proofs was presented at the \texttt{Dafny} 2025 workshop. The separation logic Iris framework was recently enhanced with heap-dependent assertions~\cite{10.1145/3729284}. The \texttt{Z3} Axiom Profiler~\cite{10.1007/978-3-030-17462-0_6} can be used to understand why certain proof obligations involving quantifiers are slow to verify. Chakarov et al. propose a method~\cite{10.1007/978-3-030-99524-9_23} to back-translate SMT counterexamples to \texttt{Dafny} syntax. \texttt{Knuckledragger}~\cite{knuckledragger2025} is an effort to turn \texttt{Z3} into an interactive theorem prover. 

\section{Discussion}

Our current implementation is a proof of concept that shows the potential of the idea, but it has several limitations, which we discuss in this section. 

The main goal of the proof mode that we propose is to be as faithful as possible to the \texttt{Dafny} verification chain, without re-implementing the verification chain itself in the interactive proof mode. Faithfulness ensures that proofs obtained at the IPM level work at the \texttt{Dafny} level as well. The main difficulty is that the \texttt{Dafny} verification chain does not support incremental compilation\footnote{Adding an incremental compilation/verification feature would require significant refactoring work on the \texttt{Dafny} internals.}, and therefore the proof mode cannot query \texttt{Dafny} itself at every proof step. For this reason, the \texttt{Dafny} IPM that we propose works by back-translating the SMT-LIB verification condition into \texttt{Dafny} syntax. In order to ensure faithfulness, we call the solver with the same configuration options that \texttt{Dafny} uses: we use it in incremental mode, we turn off the model-based quantifier instantiation option, we box integers with \texttt{LitInt}, etc. We currently concentrate on proof obligations generated with the \dafnyinline{}{}{}!{:isolate-assertions}! option and we do not support proof obligations arising from well-formedness conditions. The instrumentation code is implemented as a fork of the \texttt{Dafny} system, but in the future we plan to integrate it back into stock \texttt{Dafny} as a command line option.

The proof mode currently supports only a restricted subset of \texttt{Dafny}, including Booleans, integers, and arithmetic operations. In future work, we plan to extend this subset, starting with the functional parts and continuing with the imperative and object-oriented features. As part of this work, we will construct a database of examples for the proof mode. The prototype supports a restricted set of tactics: a cut rule (\texttt{assert}), case analysis (\texttt{case}), and assumption (\texttt{assume}). These are in a one to one correspondence with existing \texttt{Dafny} statements. We also provide a helper \texttt{check} tactic which can be used to better understand what the solver can prove automatically. In the future, we plan to expand the language of tactics to account for more complex reasoning steps, such as induction. 

We currently rely on the Python bindings for \texttt{Z3} to parse the SMT-LIB code output by Boogie. The bindings do not support commands such as \texttt{push} and \texttt{pop}. For this reason, we split the SMT-LIB file into fragments separated by \texttt{push} and \texttt{pop} using regexes and parse each fragment separately. The current prototype parses arithmetic and Boolean expressions by relying on the Python \texttt{ast} module to parse the arguments of the tactics as Python ASTs, which are then translated to \texttt{Z3} expressions. We plan to swap this for the parser generator used to parse \texttt{Dafny}. We currently only support a limited set of meta-commands, such as \texttt{quit}, \texttt{undo} and \texttt{focus}ing on a specific goal. In future work, we want to allow for additional commands for setting configuration options (such as the time limit and resource limit of the solver), searching for axioms, and others.

Another difficulty that we foresee is the back-translation of more complicated expressions. For example, a \texttt{Dafny} declaration \texttt{var x : C}, where \texttt{C} is a class, is translated into a conjunction \verb|n != null && dtype(x) == C|. We plan for the proof mode to identify these patterns and present them in \texttt{Dafny} syntax to the user. The protection functions that we use could interfere with trigger generation, and we will update the trigger generation algorithm to treat calls to \verb|_protect*| as being invisible.

To conclude, the main advantages of using the IPM over manually writing the proof directly in \texttt{Dafny} are that \( \bullet \) the full context is shown to the user at any given point, and \( \bullet \) the user can experiment with the current proof obligation without going through the potentially time consuming edit-compile-verify cycle. Finally, the IPM that we propose could be extended to modify existing proofs, could be integrated with the \texttt{Z3} axiom profiler, and be made available as a plugin to \texttt{Dafny} IDEs.

\begin{acks}
Research reported in this article was supported by an Amazon Research Award, Fall 2024.
\end{acks}

\balance
\bibliography{ref}
\end{document}